**MTF Optimization in Digital Dental X-ray Systems**

E.T.Costa and J.A.G.Albuquerque.
Departamento de Engenharia Biomédica (DEB), Faculdade de Engenharia Elétrica e de Computação (FEEC) and Centro de Engenharia Biomédica (CEB),
Universidade Estadual de Campinas (UNICAMP), Campinas, SP, Brazil

**Abstract**
In this work, we have studied the MTF optimisation relative to the detector aperture of four digital dental X-ray image systems: 1) Digora and Denoptix systems, based on PSPL detectors; 2) CDR and Sens-A-Ray 2000, based on CCD detectors. The MTF was evaluated by ERF method and linearized as a Gaussian process. The CCD based systems presented Gaussian characteristics. The PSPL based systems presented a composition of two Gaussian processes. We conclude that one process is due to the laser and stimulated light scattering inside the PSPL plate and the other is due to the laser beam focal aperture. Matching focal aperture to laser scattering allows the optimization of the PSPL systems resolution. An optimal pixel width found to be 62 μm.

**Introduction**
Digital X-ray image systems are commonly available in today general radiology. The most used direct digital X-ray transducers are the amorphous selenium (a-Se) and silicon (a-Si) photoconductive solid-state detectors, charge-coupled devices (CCD) and photo-stimulable phosphor luminescence (PSPL) detectors. Matching the aperture of the detector to its physical characteristics implies in the Modulation Transfer Function (MTF) optimisation, improving the image system resolution. We have studied MTF optimisation of four digital dental X-ray image systems: 1) Digora (Soredex, Helsink, Finland) and Denoptix (Gendex Dental System srl, Milan, Italy) systems, based on PSPL detectors; 2) CDR (Schick Technologies Inc., New York, USA) and Sens-A-Ray 2000 (Regam Medical System, Sudsvall, Sweden), based on CCD detectors.

**Method**
We have used a 0.1 mm thickness cooper edge phantom irradiated using a GE 1000 dental X-ray unit (60 kV$_p$, 10 mA e FRD = 92 cm). The MTF was evaluated by the Edge Response Function (ERF) method, as described by Xinhua et al. (2000). The 1D ERF was reconstructed from the 2D edge phantom image by averaging shifted and superimposed image columns. The column shift was given by the product of the column resolution and the sinusoidal function of the edge angle (kept between 2° and 5°). An adaptive low-pass filter was used in order to reduce noise and the Line Spread Function (LSF) was evaluated by ERF numerical differentiation. The LSF was zero padded and Fourier transformed, obtaining the MTF. Finally, MTF was normalized at zero frequency and averaged over a set of 20 images. Nickoloff and Rilley (1985) proposed a method to linearize the MTF, supposing the predominance of a Gaussian process:

$$MTF(\nu) = e^{-K\nu^2} \qquad (1)$$

where $\nu$ is the spatial frequency. Thus:

$$\sqrt{Ln(\frac{1}{MTF(\nu)})} = a\nu + b \qquad (2)$$

Note that resolution is higher as the linearized MTF slope decreases, since $a=K^{1/2}$.

**Results**

The MTF($\nu$) of the digital image systems are presented in Figure 1(a). In general, the CCD based systems presented superior resolution than the PSPL based systems. Digora system presented higher resolution than Denoptix system in 300 dpi scanning mode and lower in 600 dpi. In Table 1 we show the resolution based on the LSF Full Width at Half Maximum (FHWM) method. All resolutions are compatible with the data supplied by the manufacturers and always inferior to the Nyquist frequency. Brettle et al. (1996) carried out a study evaluating the MTF($\nu$) of the Digora system. Our data show good agreement to their results. The same authors refer that cut-off frequency (-3 dB) of Digora system is 2.5 cycles/mm. Figure 1(a) shows that CDR and SAR (CCD based systems) present similar resolution, in spite of SAR smaller aperture. The CCD based systems present superior resolution compared to that of Denoptix system (600 dpi), although presenting larger pixel size. The CDR, Digora and Denoptix (300 dpi) systems have resolutions similar to their Nyquist frequency. This indicates that resolution was heavily limited by scanning aperture. The pixel size is strongly determined by transistors dimensions, in CCD based systems, and by laser reader focal aperture, in the PSPL based systems. However, SAR and Denoptix (600 dpi) present resolution inferior to the Nyquist frequency, indicating the predominance of other limiting processes. Figure 1(b) shows the linearized MTF($\nu$) for SAR, CDR, Digora and Denoptix (300 dpi) system. Its can be noted that CCD based systems MTF($\nu$) can be approximated by a Gaussian process (R>0.996) while PSPL based systems present a deviation of this characteristic.

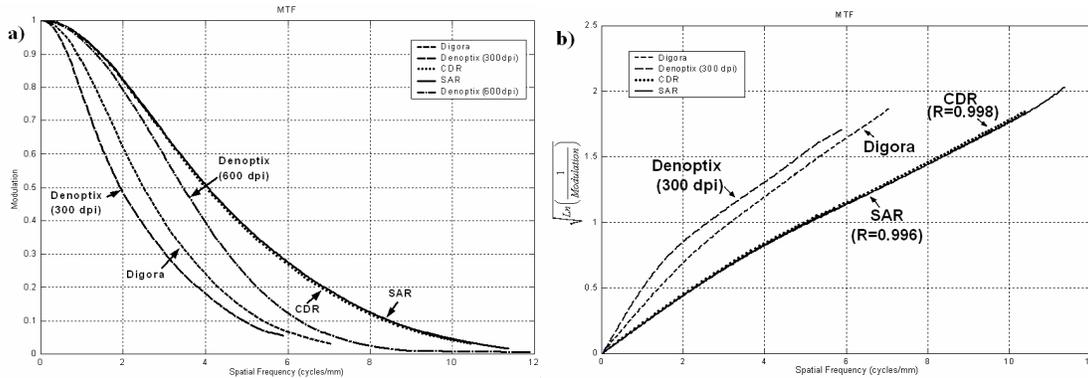

Figure 1 – (**a**) Image systems MTF($\nu$), obtained by ERF method, and (**b**) linearized MTF($\nu$).

Table 1 – Evaluated systems resolutions.

| System | Pixel width (µm) | $f_{nyquist}$ (cycles /mm) | FWHM (mm) | Resolution = 1/FWHM (cycles/mm) | Resolution by manufacturer (cycles/mm) | $f_{cut\text{-}off}$ (at -3dB) (cycles/mm) |
|---|---|---|---|---|---|---|
| CDR | 48 | 10.42 | 0.0960 | 10.42 | 10 | 4.06 |
| Sens-A-Ray 2000 | 44 | 11.36 | 0.0960 | 10.41 | 10 | 4.01 |
| Digora | 71 | 07.04 | 0.1412 | 07.08 | 6 | 2.49 |
| Denoptix (300 dpi) | 85 | 05.88 | 0.1862 | 05.37 | 6 | 1.93 |
| Denoptix (600 dpi) | 42 | 11.90 | 0.1682 | 08.56 | 9 | 3.52 |

Nevertheless, Figure 2(a) shows that the PSPL based systems also converge to a Gaussian process for spatial frequencies greater than 2.5 cycles/mm. This points to that PSPL systems

MTF($\nu$) can be linearized in parts. The MTF($\nu$) linear regression of both Digora and Denoptix systems, for frequencies above 3.5 cycles/mm, resulted in the same angular coefficient $a$=0.219 (R>0.995). This regression identifies a resolution limiting process component named process "A". Subtracting the process "A" from the original MTF($\nu$), it is obtained a second limiting process component, named process "B" and illustrated in Figure 2(b). Note that process "B" saturates near 1.7 cycles/mm.

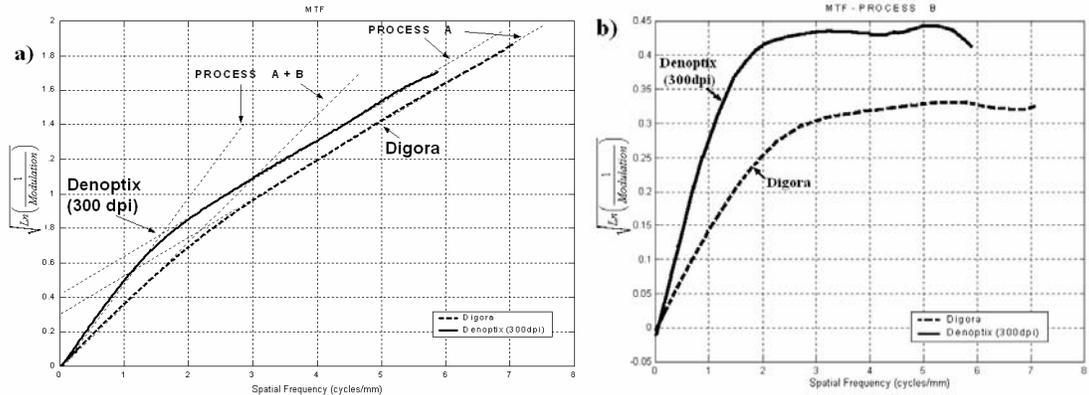

Figure 2 – **a)** Digora and Denoptix (300 dpi) image systems linearized MTF($\nu$) and **b)** process "B" MTF($\nu$) components.

Figure 3(a) shows the reconstructed MTF($\nu$) components due to process "A". Figure 3(b) shows the linearized MTF($\nu$) of Denoptix system (600 dpi). Note the saturation near 8 cycles/mm.

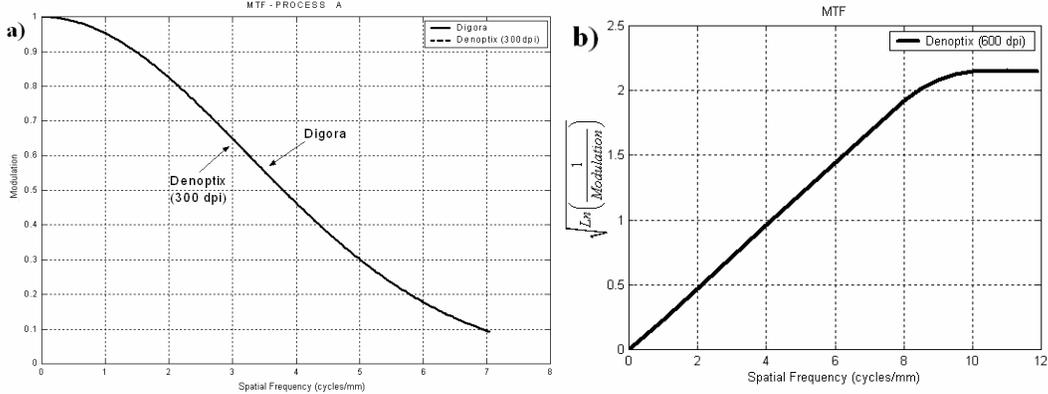

Figure 3 – **a)** Reconstructed process "A" MTF($\nu$) components of Digora and Denoptix (300 dpi) systems and **b)** linearized MTF($\nu$) of Denoptix system (600 dpi).

**Discussion**
The approximation of the CCD based systems to a single Gaussian MTF($\nu$) and agreement between resolution and Nyquist frequency indicates the predominance of the detector aperture on limiting the resolution. In other words, the X-ray unit focal spot and secondary radiation field scattering had a negligible influence in the resolution of these systems, and so in systems with larger detector apertures. We concluded that the resolution in CCD based systems is strongly determined by the transistors dimensions. However, two major processes ("A" and "B") are limiting the Digora and Denoptix (300dpi) systems. Cowen et al. (1993) discussed the main resolution limiting factors on PSPL based systems to be: 1) laser beam focal aperture of the reader; 2) laser scattering inside the phosphor storage plate; 3) stimulated light emission scattering inside the phosphor storage plate; 4) X-ray photon scattering and re-absorption inside the phosphor storage plate; 5) digital and analogical signal processing; and 6) digital sampling interval. Cowen et al. (1993) and Brettle et al. (1996) refer the laser scattering inside the PSPL plate as the main source of resolution degradation. Kashima (1995) states that both Digora and Denoptix systems PSPL plates are produced with the Fuji PSPL HR-III film. Consequently, the process "A", equal in both systems, is probably due to the laser and stimulated light scattering properties of the PSPL film. Thus, process "B" is determined by laser beam focal aperture. Figure 3(a) shows the optimal MTF($\nu$) of an image system based on this PSPL film. Therefore, it is convenient to design the laser beam focal aperture to match the PSPL film resolution. The manufacturers of Digora and Denoptix (300 dpi) selected the laser focal spot larger than the optimal pixel resolution. Denoptix (600 dpi) presents detection aperture smaller than the optimal pixel resolution, thus increasing unnecessarily the quantum noise. Quantum noise is increased by reducing the quantum fluency over samller detector element. Denoptix (600 dpi) have a MTF($\nu$) saturation in 8 cycles/mm. Digora and Denoptix systems can be optimized using a 62.5 µm pixel width (selecting the Nyquist frequency to 8 cycles/mm). In general, digital X-ray image systems must have a single Gaussian MTF($\nu$). Large pixel width will drive to a degradation process in low frequencies whereas small pixel width will drive to MTF($\nu$) saturation in high frequencies. At the same way, SAR will have better design using a 48 µm pixel width. Moreover, the scanning aperture design must take account the interest clinical diagnosis frequency range. Kashima (1995) uses clinical studies to show that useful dental clinic information is between 2 and 5 cycles/mm. According with the author, a 5 cycles/mm resolution image system would be completely capable of providing an appropriate diagnosis. Albuquerque (2001) shows that Digora system have lower Noise Power Spectrum (NPS($\nu$)) and higher Detective Quantum Efficiency (DQE($\nu$)), being the most suitable system to dental practice, instead of have lower resolution.

**Conclusions**
The method proposed in this study is convenient to isolate diverse physical process MTF($\nu$) components and optimize digital X-ray image systems design.